\documentclass[12pt]{article}
\usepackage{epsfig}
\usepackage{axodraw}
\usepackage{graphicx}
\usepackage{amsfonts,amsmath}
\usepackage{latexsym}

\def\beq{\begin{equation}}
\def\eeq{\end{equation}}
\def\bea{\begin{eqnarray}}
\def\eea{\end{eqnarray}}
\def\ba{\begin{array}}
\def\ea{\end{array}}
\def\pa{\partial}

\def\nn{\nonumber}

\def\tb{\tilde\beta}
\def\tg{\tilde\gamma}
\def\vp{\varphi}
\def\rd{{\rm d}}
\def\GeV{{\rm GeV}}

\def\cO{{\cal{O}}}
\def\cL{{\cal{L}}}

\def\hg{{\hat{g}}}
\def\hu{{\hat{u}}}
\def\hx{{\hat{x}}}
\def\hy{{\hat{y}}}
\def\hz{{\hat{z}}}

\def\LUV{{\Lambda_{\rm UV}}}
\def\LIR{{\Lambda_{\rm IR}}}
\def\CV{{V_{\rm eff}}}
\def\CW{{W_{\rm eff}}}
\def\vev{{\langle \vp \rangle}}
\def\Lev{{\langle L \rangle}}

\begin{document}

\begin{flushright}
AEI-2008-062\\LPTENS-08/50
\end{flushright}
\vspace{1cm}

\centerline{\bf\Large Renormalization Group and Effective Potential}

\vspace{3mm}
\centerline{\bf\Large in Classically Conformal Theories}

\vspace{5mm}
\centerline{\bf Krzysztof A. Meissner${}^{1}$ and Hermann Nicolai${}^2$}

\vspace{5mm}
\begin{center}
{\it ${}^1$ Institute of Theoretical Physics,
University of Warsaw,\\ Ho\.za 69, 00-681 Warsaw, Poland\\
${}^2$ Max-Planck-Institut
f\"ur Gravitationsphysik
(Albert-Einstein-Institut),\\ M\"uhlenberg 1, D-14476 Potsdam,
Germany}
\end{center}

\begin{abstract}

\footnotesize{Making use of a general formula for the RG improved 
effective (Coleman-Weinberg) potential for classically conformal models 
and applying it to several examples of physical interest, and in particular 
a model of QCD coupled via quarks to a colorless scalar field, we discuss
the range of validity of the effective potential as well as the issue 
of `large logarithms' in a way different from previous such analyses. 
Remarkably, in all examples considered, convexity of the effective 
potential is restored by the RG improvement, or otherwise the potential 
becomes unstable. In the former case, symmetry breaking becomes
unavoidable due to the appearance of an infrared barrier $\LIR$, which
hints at a so far unsuspected link between $\Lambda_{QCD}$ and the scale
of electroweak symmetry breaking.}
\end{abstract}

\noindent
{\bf 1. Introduction.} The main purpose of this paper is to clarify some
aspects of the Coleman-Weinberg (CW) mechanism of radiatively
induced breaking of symmetry \cite{CW,Jack} by means of examples,
for which the renormalization group (RG) improved versions of the
one-loop effective CW potential can be obtained in closed form. We
believe that these results constitute further evidence in support of
the scenario proposed in \cite{MN1,MN2,MN3}, according to which the
CW mechanism can be implemented in the context of the full standard
model with a classically conformal Lagrangian ({\it i.e.} without explicit
mass terms in the scalar potential), provided all couplings remain
bounded over a large range of energies. Our analysis reveals
commonly accepted wisdom concerning possible realistic applications of
the CW mechanism to be incomplete; in particular, intuition based on the
textbook examples of pure $\phi^4$ and scalar QED (which is also reviewed
here in a slightly new perspective) may fail when there are non-trivial
cancellations in the relevant  $\beta$-functions, as a consequence of
which Landau poles are shifted to very large scales. As we argued in
\cite{MN1}, this may be the ultimate reason for the stability of the
weak scale {\it vis-\`a-vis} the Planck scale $M_{Planck}$, if the
standard model couplings were to conspire in precisely such a way so as
to make the model survive to very large scales (a similar scenario, but
without radiative symmetry breaking, was proposed and elaborated in
\cite{ST,Shap}; see also \cite{Australians1,Australians2} for a
different attempt to implement the CW mechanism).

RG methods to improve the effective potential have been applied and 
studied in previous work, and for a long time; we refer readers to
\cite{IIM,CG1,CG2,Sher,Bando,Einhorn,Casas} for earlier treatments 
and comprehensive bibliographies, and \cite{Einhorn1,Canadians1,Canadians2} 
for more recent work along these lines. Making use of a general formula
for the RG improved effective potential at one loop which was first 
derived in \cite{CG2}, we here show how one can arrive at closed form 
expressions for the effective potential at one loop by means of exact 
solutions of the corresponding one-loop $\beta$-functions. In cases
where closed form solutions of the latter are not available, the potentials
can nevertheless be studied numerically in a  convenient and effective
manner by exploiting these general formulas. Closed form solutions 
of the $\beta$-function equations were presented for instance 
in \cite{BF} for a model of a scalar field coupled 
to fermions, and in \cite{Canadians3} for a simplified version of the 
standard model. All examples that we discuss support a key assertion 
of our previous work, namely that {\em the CW potential can be trusted 
in the range where the running couplings (expressed as functions of the 
classical field $\vp$) stay small.} It is this criterion which should 
be used to ascertain the consistency of the CW potential, rather than 
the widely used `rule of thumb', according to which certain products 
of the logarithm of $\vev$ and the pertinent input coupling must be small. 
The latter requirement is deficient inasmuch as it does not take into 
account possible cancellations, whereas our results show that the 
running couplings may stay small in the presence of such cancellations 
despite `large logarithms'.

Rather unexpectedly, the present work also sheds new light on another 
long-standing issue, namely the apparent clash between symmetry breaking 
and convexity of the effective potential. As is well known on general 
grounds \cite{IIM,CT,HJ,WW,BF} the effective potential must be a 
{\em convex function} of the classical scalar field, a condition that 
is generically violated by the (unimproved) one-loop expressions derived 
in quantum field theory (but see \cite{WW} for a physical interpretation 
of the imaginary terms in the effective potential resulting from 
non-convexity). Remarkably, for the classically conformal theories 
we here find that, in all examples
studied, either the convexity of the potential over its domain of
definition is restored by the RG improvement, or otherwise the
potential develops an instability~\footnote{Although these
statements seem not to apply in the presence of explicit mass terms,
when the CW potential only represents a small correction to the
classical potential, the point, in our opinion, even then requires
further study.}. The restoration of convexity is mainly due to the
fact that the effective potential not only exhibits an ultraviolet
(UV) barrier at the location $\LUV$ of the Landau pole, but in
general also an infrared (IR) barrier $\LIR$ which arises through
the couplings of the scalar field to the other fields via the coupled
system of RG equations. The allowed regions for $\vp$ are
thus separated by a `forbidden zone' $|\vp| < \LIR$.~\footnote{Of course,
 this does {\em not} mean that in a path integral treatment one should cut out
 a region around the origin in field space. The variable $\vp$ here is the 
 {\em effective classical field} defined as $\vp = \delta W/\delta J$ from 
 the generating functional $W[J]$ of connected Green's functions, and as 
 such must obviously be distinguished from the scalar field over which
 one integrates in the path integral.} In this way, the conflict between 
convexity on the one hand, and the existence of non-trivial vacua with 
$\vev\neq 0$ disappears. When the running coupling turns negative before 
this barrier is reached, the expression for the one-loop potential 
becomes unbounded from below for very small $\vp$.

In the explicit examples we shall see that the Landau pole $\LUV$
can be shifted to very large values. By contrast, in semi-realistic
models involving the strong interactions, the IR barrier $\LIR$ is
unmovable because its value is in essence dictated by the known IR
properties of $\alpha_s$. {\em If $\LIR > 0$, we must have $\vev
\neq 0$, and symmetry breaking becomes unavoidable!} To be sure, our
QCD-like example is not yet fully realistic in that the minimum is
too close to $\LIR$, whereas the scale of electroweak symmetry
breaking is more than a hundred times larger than $\Lambda_{QCD}$.
Nevertheless, we find it most remarkable how the strong interactions
-- so far not thought to play any role in this context -- may
intervene to enforce breaking of electroweak symmetry (and conformal
invariance). Readers may recall that symmetry breaking in the
standard model is conventionally implemented by means of an explicit
mass term $m^2\vp^2$ --- leaving us with the question why nature
should prefer a negative value of $m^2$ over the equally consistent
positive value!

Evidently, the poles at $\LUV$ and $\LIR$ both signal a breakdown of
perturbation theory. However, while there has been much discussion in
the literature about the Higgs coupling becoming strong in the UV, we
are not aware of similar discussions dealing with strong coupling in
the IR. The UV Landau pole is usually interpreted as the scale at which
`new physics' (here: quantum gravity) sets in. The interpretation of
the IR pole, on the other hand, is not as obvious as in QCD, where it
implies a new phase with confinement. 

For simplicity we consider only a single real scalar field with couplings
to different non-scalar fields (see e.g. \cite{Einhorn2,Bando1} for a 
discussion of the multi-field case); this has the advantage that the full
result takes a rather simple form (cf. (\ref{Effpot}), (\ref{f}), (\ref{F}) and
(\ref{Weff}) below) which is completely determined by the $\beta$-function.
The multi-field case, where such simplifications are presumably absent,
will require separate study. After explaining some general features
of the RG improvement procedure in section~2, we first consider
scalar QED in section~3, not only recovering known results, but
also to expose an IR instability that has gone unnoticed so far. In
section~4, we turn to our main example, QCD coupled to a colorless scalar
field via Yukawa interactions. This example incorporates some essential
features of the model investigated in \cite{MN1,MN3} in that the various
running couplings keep each other under control over a large range of
energies so as to ensure the survival of the theory up to some large
scale. Finally, we present some numerical results and discuss the
normalization of couplings in terms of physical parameters.

\vspace{3mm}\noindent
{\bf 2. Generalities.} In a classically conformally invariant
theory~\footnote{See e.g. the recent article \cite{K} for a comprehensive
review of conformal invariance in field theory and quantum field theory.}
with one real scalar field that couples to any number of fermions and/or
gauge fields, the effective CW potential is generally of the
form~\footnote{We generally write $\CW$ for the RG improved, and
$\CV$ for the unimproved effective potential at one loop.}
\beq\label{Effpot}
\CW(\vp) = \vp^4 f(L,g)
\eeq
where
\beq\label{L}
L\equiv \ln \frac{\vp^2}{v^2}
\eeq
and $v$ is some mass scale required by regularization. The letter $g$
in (\ref{Effpot}) stands for a collection of coupling constants
$\{g_1,g_2,\dots\}$ corresponding to the quartic scalar self-coupling
and various (dimensionless) couplings to and among other fields
(fermions, gauge fields) in the theory.  Below we will also use letters
$u,x,y,z$ from the end of the alphabet  to denote convenient combinations
of these couplings, as they occur in the $\beta$-functions.

The quantity $v$ is the only dimensionful parameter in the theory, but
has no physical significance in itself; eventually, it should thus be
replaced by a more physical parameter, such as the vacuum expectation
value $\vev$. When one uses dimensional regularization (as we do),
$v$ enters via the replacement
\beq
\int \frac{d^4 k}{(2\pi)^4} \;\; \longrightarrow \;\;
v^{2\epsilon}
\int \frac{d^{4-2\epsilon} k}{(2\pi)^{4-2\epsilon}}
\eeq
which is to be performed in all divergent integrals. The parameter
$v$ breaks conformal invariance explicitly, and the question is then
whether this breaking persists after renormalization, in which case
the classical conformal symmetry is broken by anomalies. For the
renormalization we employ the prescription of \cite{MN2}, according
to which the local part of the effective action is to be kept conformally
invariant throughout the regularization procedure: in this way, the
structure of the anomalous Ward identity is preserved at every step
(as originally suggested in \cite{Bardeen}). As a consequence, the mass
parameter $v$ appears in the effective action only via (\ref{L}), 
{\it i.e.} logarithmically.

The function $f$ must be determined from a perturbative expansion
\cite{CW,Jack}. A main problem then is to assess the reliability of
such approximations, and to ascertain whether extrema found by minimizing
the perturbative potential are within the perturbative range or not.
On general grounds, the effective potential $\CW$ must satisfy
the RG equation (see e.g. \cite{IIM,PR,Sher})
\beq\label{RGW}
\left[ v\frac{\pa}{\pa v} + \sum_j \beta_j (g)\frac{\pa}{\pa g_j}
  + \gamma(g) \vp \frac{\pa}{\pa\vp}\right] \CW(\vp,g,v) = 0
\eeq
where $\beta_j(g) \equiv \beta_j(g_1,g_2,\dots)$ and
$\gamma(g) \equiv \gamma (g_1, g_2, \dots)$ are the relevant
$\beta$-functions and anomalous dimension, respectively; they depend on
the theory under consideration. Substituting (\ref{Effpot}) into this 
formula, we obtain \cite{CG2}
\beq\label{RG1}
\left[-2\frac{\pa}{\pa L} +
 \sum_j \tb_j (g)\frac{\pa}{\pa g_j}+4\tg(g)\right]f(L,g_j)=0
\eeq
where
\beq
\tb(g) \equiv \frac{\beta(g)}{1 - \gamma(g)} \quad , \qquad
\tg(g) \equiv \frac{\gamma(g)}{1 - \gamma(g)}
\eeq
We note that the effect of the rescaling by the factor $(1-\gamma)$ will
only appear in higher orders, and thus not play any role for the (one-loop)
considerations in this paper. For this reason, we will omit the tildes
in all formulas in the following sections. A perturbative analysis of these
equations was already begun in \cite{Canadians1,Canadians2}, and a closed
form of the solution of the one-loop RG equations, albeit very complicated,
was derived for a simplified version of the standard model in
\cite{Canadians3}.

The general solution of (\ref{RG1}) (to {\em any} order) is obtained  by first
solving the system of ordinary differential equations for the running
couplings, viz.~\footnote{For notational clarity we always put hats
on the $L$-dependent running couplings.}
\beq\label{RG2}
2\frac{\rd}{\rd L} \hat{g}_j (L)
   =   \tb_j\big(\hat{g} (L)\big)
\eeq
where the initial values $\hat{g}_j\big(0\big)$ are the input parameters
from the classical Lagrangian. Given a solution (\ref{RG2}), the partial
differential equation (\ref{RG1}) is solved by
\beq\label{f}
f(L,g_j) \equiv F\big(\hg_1(L), \hg_2(L), \dots \big)
\exp \left[ 2\int_0^L \tg(\hg (t)) \rd t \right]
\eeq
where $F$ is an a priori {\em arbitrary} function, which can be
determined by matching it with the perturbative (loop) expansion
of the effective action. The precise choice of $F$ together with
the choice of $\beta$-functions fixes the renormalization scheme
(recall that the $\beta$-functions are renormalization scheme dependent
in higher loop order). We take
\beq\label{F}
F\big(\hat{g}(L)\big) \equiv \hg_1(L)
\eeq
where $g_1$ is the scalar self-coupling. In this way we arrive
at the general formula
\beq\label{Weff}
\CW(\vp,g,v) = \hg_1(L) \vp^4 
\exp\left[ 2\int_0^L \tg(\hg (t) \rd t \right]
\eeq
At any order in the loop expansion, (\ref{F})  is the most
natural choice because the effective potential (\ref{Weff}) evidently
reduces to the classical potential in the limit $\hbar\rightarrow 0$
(we recall that each factor of $L$ is accompanied by a factor $\hbar$, 
but we usually set $\hbar =1$). Other schemes can
differ from this one by higher powers of the (running) coupling
constants on the r.h.s. of (\ref{F}), such that
\beq\label{F1}
F\big(\hat{g}(L)\big) = \hat{g_1}(L) +
\sum_{i,j} \alpha_{ij} \hat{g}_i(L) \hat{g}_j(L) + \dots
\eeq
corresponding to a non-linear redefinition of the coupling
constants. We note that whenever closed form solutions of (\ref{RG2}) 
can be found, (\ref{Weff}) yields a {\em completely explicit} formula 
for the RG improved potential. Morever, even if such explicit solutions
do not exist, (\ref{Weff}) can be conveniently exploited to explore
the RG improved potential numerically (as we do in section~4).

Obviously the above expression for the RG improved effective potential 
(and (\ref{Weff}) in particular) can be trusted as long as the running 
couplings $\hg_j(L)$ remain small (in which case the higher order corrections
in (\ref{F1}) are likewise small). Imposing smallness of the running
couplings in turn determines an admissible range of values for $L$,
and thereby for the field $\vp$ itself. In realistic applications we
will seek to ensure by judicious choice of the couplings that this
admissible range extends beyond the Planck mass for positive $L$,
but we usually will also encounter a {\em lower} bound for negative
$L$. With the replacement of the one-loop effective potentials by RG
improved potentials, there is no more need to consider running
coupling constants. Rather, the energy scale is now replaced by the
classical field $\vp$ itself, or more precisely, the ratio $\vp/v$.
Let us also recall that the one-loop RG improved effective action
corresponds to a resummation of the perturbation series where only
leading logarithms are kept.

To explain the main point in a nutshell, let us consider the textbook
example of pure ({\it i.e.} massless) $\phi^4$ theory, for which the
one-loop effective potential is \cite{CW}
\beq\label{CW1}
\CV(\vp)= \frac{g}4 \vp^4 + \frac{9g^2}{64\pi^2} \vp^4 L
\eeq
The RG improved potential is found by first solving the RG equation
for the scalar self-coupling
\beq
 2 \frac{\rd\hy}{\rd L} = \frac92 \hy^2 \qquad \mbox{with} \;\;\;
\hy(L)\equiv \frac{\hg(L)}{4\pi^2}
\eeq
The well known solution is \cite{PR}
\beq
\hy(L) = \frac{y_0}{1-(9/4) y_0 L}
\eeq
exhibiting the famous Landau pole at $\vp = \LUV \equiv v
\exp[2/(9y_0)]$. From (\ref{Weff}), the RG improved
potential at one loop is therefore given by
\beq\label{CW2}
\CW (\vp, g_0) =
\frac{\pi^2 y_0 \vp^4}{1 - (9/4) y_0L} = \pi^2 y_0\vp^4
   \Big[1 + (9/4) y_0 L + \dots \Big]
\eeq
(the anomalous scaling can be neglected because $\gamma(g)=0$ at one
loop). This expression exhibits the very same Landau pole as the scalar
self-coupling, and in this way limits the range of trustability to
the values $y_0L < 4/9$, or equivalently $|\vp |< \LUV$ (there is no IR
barrier, so $\vp$ can become arbitrarily small). Because there is no
non-trivial minimum of $\CW$ in this range we recover the well-known
result that the symmetry breaking minimum in the unimproved potential
$\CV$ (\ref{CW1}) is spurious. The existence of a spurious minimum $\vev
\neq 0$ in (\ref{CW1}) is related to the fact that the function (\ref{CW1})
is {\em not convex}. The RG improved version (\ref{CW2}), on the
other hand, {\em is} convex, and therefore the minimum is moved back
to $\Lev = -\infty$, that is, $\vev = 0$. As we will see in our
`real world' example (and also in scalar QED), the latter
possibility is precluded by an IR barrier, which entails $\Lev > -
\infty$.

\vspace{3mm}\noindent{\bf 3. Massless scalar QED revisited.} Next we turn
to massless scalar QED, which describes the coupling of one complex
scalar field to electromagnetism. Although this is generally to be
well understood, and was discussed in detail in \cite{CW} we would
like here to emphasize one specific feature of the model, namely a
potential IR instability. Introducing
\beq
y=\frac{g}{4\pi^2},\ \ \ \ u=\frac{e^2}{4\pi^2}
\eeq
for the scalar self-coupling and the electromagnetic coupling,
respectively, we have the system of equations \cite{CW}
\beq\label{RGQED}
2\frac{\rd \hy}{\rd L} =  a_1\hy^2-  a_2\hy\hu+a_3 \hu^2\quad ,
\qquad 2\frac{\rd \hu}{\rd L} =   2b\hu^2
\eeq
where for the scalar QED $a_1=\frac56,\ a_2=3,\ a_3=9$ and
$b=\frac1{12}$. At this order, we also have
\beq
\gamma (y,u) = cu
\eeq
with $c=\frac34$. The solution was found in \cite{CW}
\bea\label{uyQED}
\hu(L)&=&\frac{u_0}{1-b u_0L}\nn\\
\hy(L)&=&\frac{(a_2+2b)\hu(L)}{2a_1}+\frac{A\hu(L)}{2a_1}
\tan\left(\frac{A}{8b}(\ln(\hu(L))+C)\right)
\eea
with (we assume that the constants give $A>0$ as is the case for
scalar QED)
\beq\label{A}
A=\sqrt{4a_1a_3-(a_2+2b)^2}
\eeq
and $C$ should be chosen to satisfy $\hy(0)=y_0$.

As we explained, $\hy(L,y_0,u_0)$ then also solves the RG equation (\ref{RG1})
\beq\label{QED1}
\left[-2\frac{\pa}{\pa L}+\beta_y\frac{\pa}{\pa y}
+\beta_u\frac{\pa}{\pa u}\right]F(L,y,u)=0
\eeq
with the proper classical limit $F(0,y,u)=y$. Hence, from (\ref{Weff})
we obtain the {\em exact} RG improved effective potential
at one loop
\beq\label{VQED}
\CW (\vp,y,u) =
\pi^2 \vp^4 \, \hy(L)/(1-b u_0L)^{2c/b}
\eeq
Let us first recall the standard treatment of (\ref{VQED}), as explained
in \cite{CW}. By taking $y$ and $u$ small, with the additional relation
$y=\alpha u^2$, and $\alpha=\cO(1)$, one easily checks
that the minimum occurs at $\Lev =O(1)$ independently of the input value
of the coupling $u$. Because $u\Lev$
is thus small, one concludes that the symmetry breaking minimum
is not spurious, unlike for pure $\phi^4$.

Inspection of the RG improved potential (\ref{VQED}) reveals 
one feature of the full potential which is not visible
in the unimproved potential. Because of the tangent
function in the solution there are now {\em two} barriers -- one is
the usual UV Landau pole, whereas the other represents an IR barrier,
disallowing small values of $\vp$. There is, however, an important
difference between the two --- at the IR barrier the potential
becomes unbounded from below. Hence, though inside the region of
applicability of perturbation theory, the symmetry breaking vacuum
is at best {\em meta-stable}. This clearly indicates a breakdown of
perturbation theory in the IR; although it is conceivable that
higher orders might re-stabilize the effective potential there,
one will have to resort to non-perturbative methods in order
to settle the question.

The unboundedness of  $\CW$ in the IR can be traced back the strict
positivity of $\hy' (L)$ in (\ref{RGQED}).
The IR instability can be avoided by pushing $\ln(\LIR/v)$ back to 
$-\infty$, as may happen for other choices of the input parameters, but 
then we are back to (\ref{CW2}) and there would be no symmetry breaking. 
When the RG improved potential $\CW$ is unstable, it also fails to be 
convex, exemplifying the claimed link between instability and lack of
convexity.

\vspace{3mm}\noindent {\bf 4. QCD with one colorless scalar.}
Our main example is closely modeled on \cite{MN1,MN3}, and thus
incorporates features of the Standard Model. The Lagrangian can be written
schematically as
\beq\label{Lag}
\cL = - \frac14 {\rm Tr}\, F_{\mu\nu} F^{\mu\nu} +
      \bar{q}^i \gamma^\mu D_\mu q^i -
      \frac12 \partial_\mu\phi \partial^\mu \phi +
      g_Y \phi \bar{q}^i q^i - \frac{g}4 \phi^4\,.
\eeq
It is classically conformally invariant and depends on three couplings,
the gauge coupling $g_s$, the Yukawa coupling $g_Y$ and the scalar
self-coupling $g$. The (real) scalar field $\phi$ is not charged under
Yang-Mills $SU(N)$ (hence colorless), but couples to color-charged quarks
via the Yukawa term, much like the standard model Higgs. The main advantage
of the model (\ref{Lag}) is that the one-loop RG equations can again be solved
exactly.  This enables us to exhibit new phenomena that are not visible in
the usual perturbative expansion, and which require a minimum of {\em three}
independent couplings. An important feature is that the one-loop
$\beta$-functions for the scalar and for the fermions both have two
contributions of opposite sign, so that both couplings can remain stationary
over a large range of energies with suitable initial conditions.

The one-loop $\beta$-function equations are now given by the
system
\beq\label{RGQCD}
2\frac{\rd\hy}{\rd L} =  a_1\hy^2 + a_2\hx \hy-a_3\hx^2 \;\; , \quad
2\frac{\rd\hx}{\rd L} =  b_1\hx^2 - b_2\hx\hz \;\; , \quad
2\frac{\rd\hz}{\rd L} = - c\hz^2
\eeq
for the functions $\hy\equiv \hy(L)\,,\, \hx\equiv \hx(L)$ and
$\hz\equiv \hz(L)$, where
\beq
x\equiv \frac{g_Y^2}{4\pi^2} \;\; ,  \quad y\equiv \frac{g}{4\pi^2}
\;\; , \quad z\equiv \frac{g_s^2}{4\pi^2} \equiv \frac{\alpha_s}\pi
\eeq
and the anomalous dimension of $\phi$ is equal to $\gamma(x,y,z) =-ax$
at this order.

For the Standard Model the values are \cite{Sher}
\beq\label{valSM}
a_1=6,\ \ a_2=3,\ \ a_3=\frac32,\ \ b_1=\frac94,\ \ b_2=4,\ \
c=\frac72,\ \ a=\frac34
\eeq
and they are similar also to the model with massive neutrinos such
as the one in \cite{MN1}. The whole system of equations admits a
stable solution with suitable initial values, because asymptotic
freedom keeps the YM coupling under control, which in turn slows
down the running of the Yukawa coupling, which in turn keeps the
scalar self-coupling under control. This  cascade of mutual
cancellations is the mechanism invoked in \cite{MN1, MN2} to avoid
Landau poles or instabilities between the small scales and the
Planck scale. Furthermore, the model shows that indeed the one-loop
solution of RG equations for the effective potential has a minimum
for a field much below the value of the field where the UV Landau
pole occurs and the crucial point that we want to emphasize in this
paper is that there exists an IR Landau pole so that the resulting
effective potential is simultaneously convex and leads to symmetry
breaking.

In the case of the Standard Model a closed form solution of (\ref{RGQCD})
was presented in \cite{Canadians3}, with a (complicated) ratio of
hypergeometric functions as the result. The generic feature of the 
solution is the presence of either a Landau pole or an instability 
on the UV side and a Landau pole on the IR side (caused by the strong 
coupling evolution), depending on the initial values of the coupling 
constants. Since the explicit expressions are not very illuminating
let us here concentrate on the behavior of the solution near the
IR Landau pole, which is the most interesting feature of this model.
The solutions for $\hz(L)$ and $\hx(L)$ can be easily found
\bea
 \hz(L)&=&\frac{z_0}{1+c z_0 L/2},\ \ \nn\\
 \hx(L)&=&\frac{(b_2-c)\hz(L)}{b_1-K \hz(L)^{1-b_2/c}}
\eea
where $K$ should be chosen to satisfy $\hx(0)=x_0$. We also need
the formula
\beq
\int_0^L \hx(t)\rd t=-2\ln\left[
\frac{b_1 \hz(L)^{b_2/c-1}-K}{b_1 z_0^{b_2/c-1}-K}\right]
\eeq
For  $b_2>c$ (as is the case for (\ref{valSM})) the power of $\hz(L)$
in the denominator of the expression for $\hx(L)$ is negative so
independently of the initial value $\hx(0)$
\beq
\hx(L)\to \beta \hz(L),\ \ \ \ \beta=\frac{b_2-c}{b_1}
\eeq
when $L\to\ln(\LIR^2/v^2)$ (i.e. $\hz(L)\to \infty$). The IR pole is
obviously at
\beq
\LIR=\exp\left(-\frac{1}{c z_0}\right)
\eeq
Plugging these solutions into the equations for $\hy(L)$ we get in the
same limit
\beq
\hy(L)\to \alpha \hz(L), \ \ \ \alpha=\frac{ - a_2\beta-c+\sqrt {
(a_2\beta+ c)^2+4 a_1 a_3 \beta^2}} {2a_1 }
\eeq
The effective potential has a correction due to the anomalous
dimension; in the same limit it reads
\beq
\CW(L)\to \left(\frac{b_1 \hz(L)^{b_2/c-1}-K}{b_1
z_0^{b_2/c-1}-K}\right)^{2a}\alpha \hz(L)
\eeq
Therefore the IR pole is a generic feature of this quasi-realistic
system. The function $W(L)$ is convex due to the presence of two poles,
and the symmetry breaking is unavoidable.

\vspace{3mm}\noindent{\bf 5. Numerics and normalization.}
As the solution of (\ref{RGQCD}) is somewhat cumbersome, we have
investigated it numerically for values (\ref{valSM}) given by the
Standard Model and a variety of values of the input parameters. In
all cases, we find that the solution is either convex or unstable,
in agreement with our general claim. Concentrating on the first case
for its obvious physical interest a typical set of values is
\beq\label{num1}
x_0 = 0.120 \;\; , \quad y_0 = 0.020 \;\; , \quad z_0 = 0.249
\eeq
For these parameters, we display the running coupling $\hy(L)$ and
the effective potential $\CW$ in Fig.~1 and Fig.~2, respectively;
note that the scales for $L$ in the two figures are very different.
As one can see, $\hy(L)$ stays well behaved up to very large values
$L \sim 100$. We thus see that the smallness of $\hy(L)$ does not
necessarily require the product $6\hy(0)L$ to be small (near the
Landau pole, we have $6\hy(0)L \sim 15$), showing that the
approximation can be trusted in spite of `large logarithms'. On the
IR side, we have a pole at $\ln (\LIR/v) \sim - 2.29$, while the
minimum is located at $\Lev \sim - 1.73$, where $\hg(\Lev) = 0.003$.
This `closeness' of the minimum and the IR barrier is the only
non-realistic feature of our model. Our curves not only put in
evidence the convexity of $\CW$ but also show that $\hg_1(L)$ is
very flat over a large range of values for $L$.

Let us also comment on the question of how to normalize the couplings.
Our choice (\ref{F}) corresponds to normalizing all couplings at a fixed
but arbitrary value $\vp=v$ where $L=0$, with the values $\hg_j(0)$ as
the input parameters from the classical Lagrangian. However, because the
normalization parameter $v$ has no physical significance in itself, and
having found a non-trivial minimum at $\vp =\vev$, one would like to change
this normalization {\it a posteriori} to one where $v$ is traded for the
actual value of $\vp$ at the minimum (as is usually done for $\phi^4$
and scalar QED \cite{CW}). This is desirable in view of the fact that all 
physical quantities (masses, effective couplings) are obtained as derivatives
of the effective potential at the minimum.~\footnote{Clearly, it makes
no sense to normalize the couplings at $\vp =0$, not only because
the fourth derivative diverges at $\vp =0$ for the unimproved effective
potential $\CV$, but also because this value is outside the domain of
definition of $\CW$ if there are IR barriers.}

Quite generally, having determined $\vev$, we can evolve $\hg_1(L)$ from
$\hg_1(0)$ to $\hg_1(\Lev)$, and try to fix the latter `backwards'
to some given value by varying the input parameters. Alternatively, we
can define the scalar self-coupling as the fourth derivative of $\CW$
at the minimum. Thanks to (\ref{Effpot}), (\ref{f}) and (\ref{F}), we 
can work out the relation between these two quantities explicitly.
Neglecting the contributions from the anomalous rescaling, we get
\bea
\!\!\!\!\!\!\!\!
\frac1{24\pi^2} \frac{\rd^4 F (\vp,L)}{\rd \vp^4} \Bigg|_{\vp = \vev}
 &=& \hg_1(\Lev) \, + \, \Bigg\{ \frac{25}6 \frac{\rd \hg_1(L)}{\rd L}
  \,+\, \frac{35}6 \frac{\rd^2 \hg_1(L)}{\rd^2 L} \, +    \nn\\
 && \qquad + \, \frac{10}3 \frac{\rd^3 \hg_1(L)}{\rd^3 L} \, +\,
   \frac23 \frac{\rd^4 \hg_1(L)}{\rd^4 L} \Bigg\} \Bigg|_{L = \Lev}
\eea
Because, at the minimum, we have $(2\hg_1 + \hg_1')|_{L=\Lev} = 0$, the
difference could be appreciable, but there are cancellations since
$\hg_1' <0$ while $\hg_1'' > 4 \hg_1$ by convexity. Indeed, with the
values (\ref{num1}) we find that the l.h.s $\sim 0.156$,
while $\hg_1(\Lev)= 0.196$. We have checked for a range of input couplings
that the difference between the two numbers does remain rather small.

Similar comments apply to the other couplings. Taking the Yukawa interaction
as an example, the RG improved version of the corresponding term in the
effective action takes the form
\beq
\Gamma_Y(\vp,q,\bar{q}) = h(L) \vp \bar{q} q
\eeq
where
\beq
h(L) \equiv \hg_Y(L) \exp\left[ \int_0^L \left( \gamma(t) +
2\gamma_q(t)\right) dt \right]
\eeq
As before, we can evolve this coupling from $\hg_Y(0)$ to $\hg_Y(\Lev)$
and compare with the relevant derivative. This gives
\bea
m_q &=& \vev \, \hg_Y(\Lev) \nn\\
\frac{\rd}{\rd\vp} \Big( \vp h(L) \Big) \Big|_{\vp = \vev}
    &=& \hg_Y(\Lev) + 2\frac{\rd \hg_Y(L)}{\rd\vp} \Bigg|_{L=\Lev}
\eea
respectively, for the fermion mass and the effective Yukawa coupling
at the minimum. Again, we see that the difference between the latter
and $\hg_Y(\Lev)$ can be small provided $\hg'_Y$ is small there. These
considerations justify (to some extent) the approximation used \cite{MN1,MN3},
where we effectively defined the couplings in terms of derivatives at the
minimum of $\CV$, and took those values as the input parameters
at $\vp = \vev$ to analyze the evolution of couplings.

\vspace{3mm}\noindent{\bf 6. Discussion.}
Much of our discussion was based on exact solutions of the
$\beta$-function equations, but the extension of the present considerations
to more realistic examples is straightforward. In particular, the
standard model with one Higgs doublet falls into the class of models
investigated here, since there remains only one real scalar field after
absorption of three scalar degrees of freedom into the massive vector
vector bosons. Even when closed form solutions of the RG equations are
no longer available, we can solve numerically for the running couplings 
and determine the RG improved effective potential from the general
formulas (\ref{Effpot}) and (\ref{F}). The latter can be analyzed numerically
along the lines described here. As already pointed out, the only non-realistic
feature here is the closeness of the minimum to the IR barrier $\LIR$,
which appears to be a generic feature of the one-field case. In realistic
applications, on the other hand, we would have to arrange
$\LIR\sim\Lambda_{QCD}= \cO(1 \,\GeV)$ and $\vev = \cO( 200 \, \GeV)$.
This may indicate the need for extra scalar fields (which are anyhow needed
for the inclusion of massive neutrinos \cite{ST,MN1,Shap,MN3,Australians1}).

We believe that the present analysis strengthens the case for the
applicability of the CW mechanism in a realistic context, if Landau
poles can be shifted beyond the Planck scale. In addition, it puts
in evidence the potential importance of IR poles in the scalar sector 
of the standard model, which may arise through the coupled RG equations. 
The phenomenon of the Higgs coupling becoming strong in the IR appears
puzzling, and its physical consequences remain to be explored. Somewhat 
ironically, our analysis also shows that scalar QED, often cited as the 
showcase example of the CW mechanism, suffers from a potential IR instability 
(or at least from an IR breadown of perturbation theory). It would be 
interesting to work out the consequences of the present results for cosmology,
and in particular for models of scalar field inflation, where the effective
potential (rather than the classical potential) should play a
decisive role. We would not be surprised if our results can be used
to rule out many of the currently popular `designer potentials' for
inflation.

\vspace{0.5cm}
\noindent{\bf {Acknowledgments:}} We are grateful to E.~Weinberg and
P.~Chankowski for helpful comments and criticism on a first version of 
this paper. H.N. thanks \'Ecole Normale Sup\'erieure, Paris and 
B.~Julia for hospitality and support while part of this research
was being carried out. K.A.M. was partially supported by the
EU grant MRTN-CT-2006-035863 and the Polish grant N202 081 32/1844,
and thanks AEI for hospitality and support during this work.


\newpage
\vspace{1cm}
\begin{center}
\includegraphics[width=7.5cm,viewport= 5 15 550 540,clip]{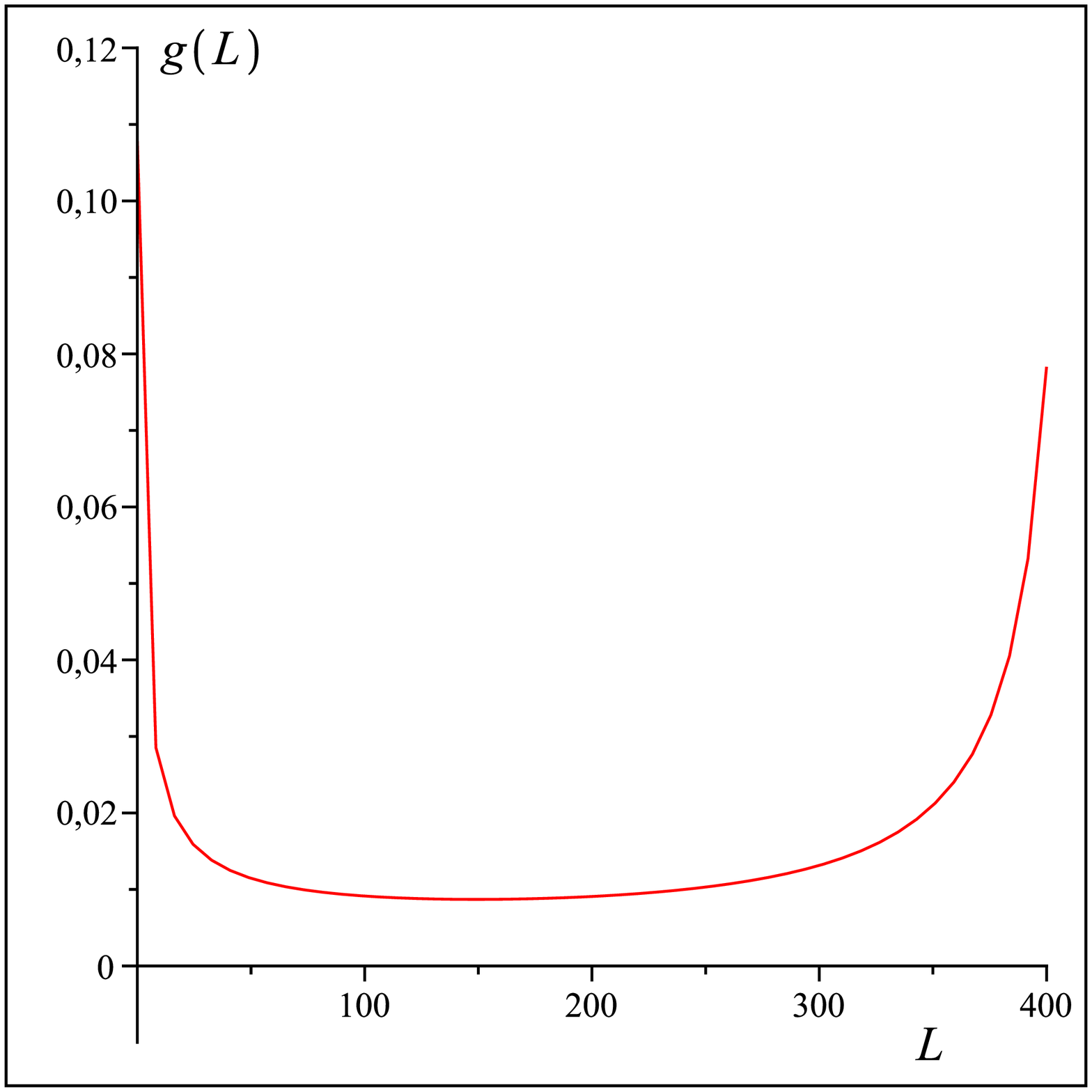}

\vspace{5mm}
{\it Fig. 1. The scalar self-coupling for the model (\ref{Lag}) with 
             (\ref{num1}).}

\vspace{2cm}
\includegraphics[width=7.5cm,viewport= 5 15 550 540,clip]{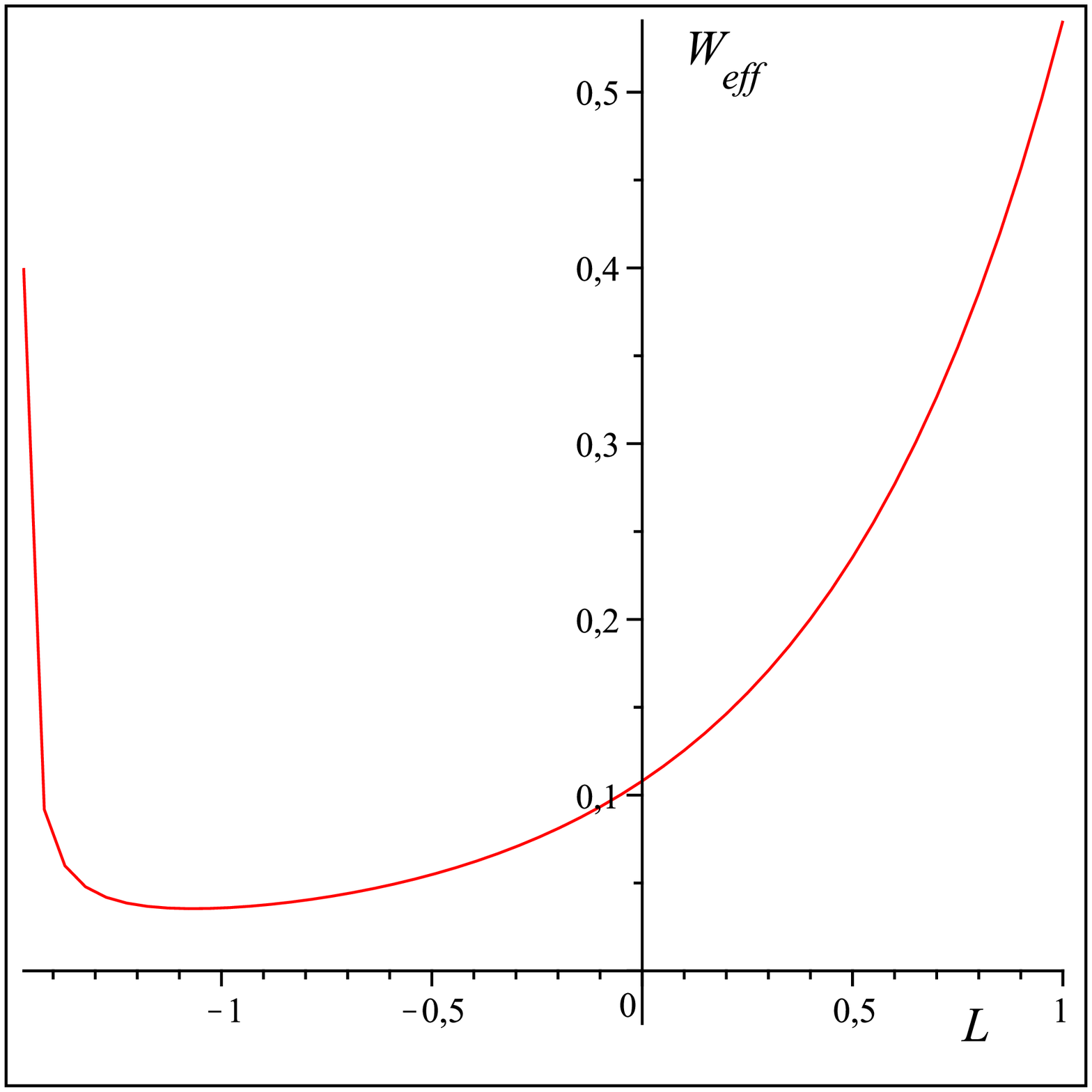}

\vspace{5mm}
{\it Fig. 2. The RG improved effective potential for (\ref{Lag})
             with (\ref{num1}).}
\end{center}

\end{document}